# Switchyard design for the Shanghai soft X-ray FEL facility

Duan Gu[1]    Zhen Wang[1]    Da-Zhang Huang[1]    Qiang Gu[1]    Meng Zhang[1,*]

**Affiliations:**

[1]Shanghai Institute of Applied Physics, Chinese Academy of Sciences, Shanghai 201800, China

[*]zhangmeng@sinap.ac.cn

**Abstract:** In this article, a feasible switchyard design is proposed for the Shanghai soft x-ray free electron laser facility. In the proposed scheme, a switchyard is used to transport the electron beam to different undulator lines. Three-dimensional start-to-end simulations have been carried out to research the beam dynamic during transportation. The results show that the emittance of the electron beam increases less than 10%, at meanwhile, the peak current, the energy spread and the bunch length are not spoiled as the beam passes through the switchyard. The microbunching instability of the beam and the jitter of the linear accelerator (linac) are analyzed as well.

**Key words:** Switchyard, resistive wall effects, microbunching instability, jitter

## 1  Introduction

With the great potential of providing high-power coherent short wavelength radiation, Free electron lasers (FEL) [1] has experienced impressive development in the past decades [2]. Nowadays, most of the major accelerator labs in the world have constructed x-ray FEL facilities, going back to FLASH [3] at DESY, LCLS [4] at SLAC, SACLA [5] at RIKEN, FERMI [6, 7] at Elettra, PAL-FEL [8] at Pohang Accelerator Lab, and further facilities are constructing or upgrading, e.g. SXFEL [9] at SINAP, Swiss-FEL [10] at PSI, European-FEL [11] at DESY and LCLS-II [12] at SLAC. These facilities have offered novel experimental capabilities and applications in diverse areas of science.

However, comparing with the synchrotron radiation light source, the single-pass FEL can only support a few users to operate the experiments simultaneously. In order to explore the applications of FELs, improve the facility capabilities and meet the continuously growing user demands, building more undulator lines is a good choice for FEL facilities. Thus, a switchyard [13] is necessary to distribute the electron beam for different undulator lines. The switchyard has to be designed to guarantee that the beam quality is maintained as the beam passing through.

In this article, we present a feasible switchyard design for the Shanghai soft x-ray free electron laser facility (SXFEL). The SXFEL facility, as shown in **Fig. 1**, is a user facility upgraded from the SXFEL test facility. The linac of the SXFEL consists of an S-band photo-injector, a laser heater system, an X-band linearizer, two bunch compressors and the C-band main accelerators. There are two undulator lines, one is based on SASE scheme and the other is for cascaded seeded FEL scheme. The switchyard inserted downstream of the linac is used to transport the electron beam to the cascaded seeded FEL undulator line. The main parameters of the electron beam at the end of the linac are listed in **Table 1**.

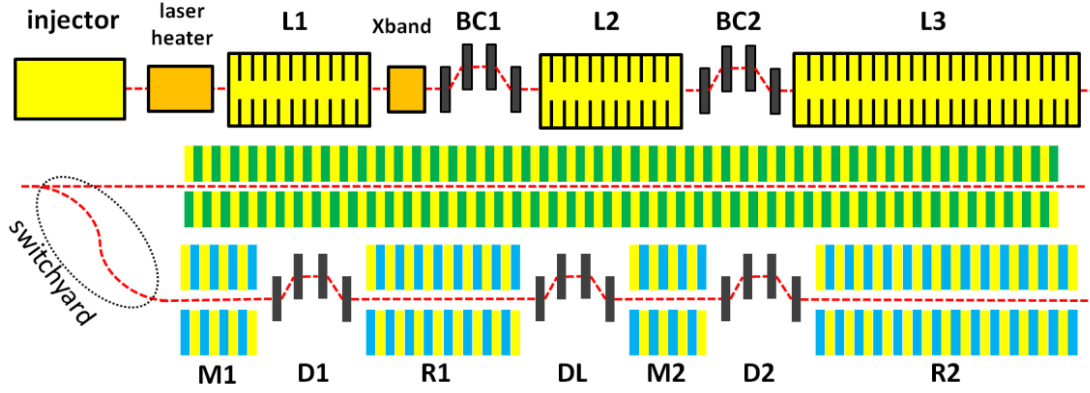

**Fig. 1** The schematic layout of the SXFEL facility

**Table 1** Main linac parameters of SXFEL

| parameter | specification | unit |
|---|---|---|
| electron beam energy | 1.5 | GeV |
| peak current | 700 | A |
| charge | 500 | pC |
| bunch length (FWHM) | ~700 | fs |
| transverse normalized emittance | 1 | mm·mrad |
| repeat frequency | 50 | Hz |

This article is organized as follows. We firstly introduce the schematic layout of the switchyard in Sec. 2. The beam collective effects, including the resistive wall effects and the micro-bunching instability effects, are shown in Sec. 3. Then in Sec. 4, we have a research on the longitudinal and transverse jitters of the electron beam. Finally, summary and conclusion are given in Sec. 5.

## 2  Switchyard description

The switchyard of SXFEL is located downstream of the linac, beginning with a fast kicker to distribute electron bunches either towards the SASE line or the cascaded HGHG line. While a 130m-long drift which originally used for undulator sector of SXFEL test facility, will be removed and replaced with a beta-matching FODO structure. In order to separate the beam, a septum will be used together with the kicker to achieve a balance for different technical difficulties on view of beam stabilities. A fast kicker followed by the septum is located at the entrance of the switchyard, aiming to deviate the target electron beam off from the straightforward orbit with an angle of 0.5 degree. The septum magnet is used to further deviate the target electron beam with an angle of 2.5 degree, while has little influence on the straightforward beam. The distance between the kicker and the septum is 1 m. There is a 3-degree bending magnet located 5 m downstream of the septum, comprising the TBA section. A quadrupole is inserted between the septum and the bending magnet to adjust the value of R56, in order to make the TBA section isochronous (so is the DBA section). The switchyard has a triple-bend (TBA) and a double-bend (DBA) achromat module, and each module has a total bending angle of 6 degrees. The TBA and the DBA section are connected with a symmetrical beta-matching FODO section, making the switchyard a total length of 40 m. The schematic layout of the switchyard is shown in **Fig. 2**.

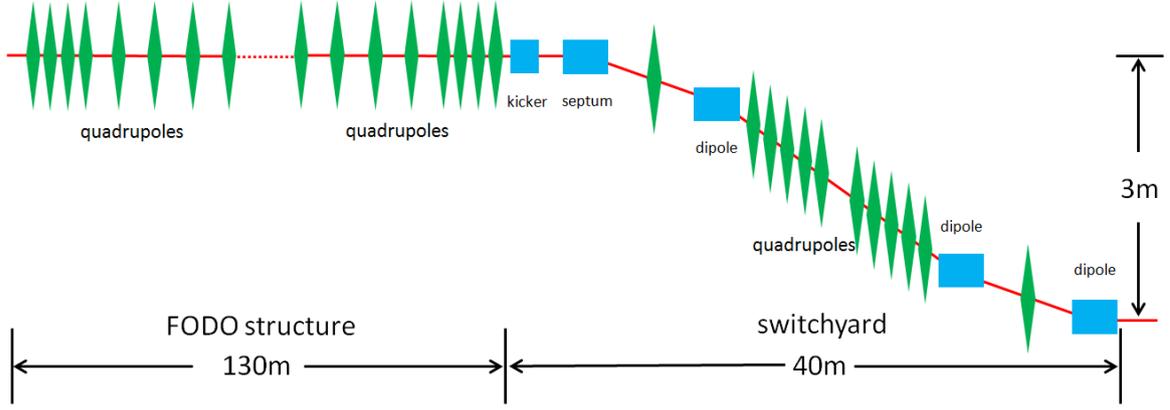

Fig. 2 The schematic layout of the switchyard

The phase advance between the TBA and DBA is designed to be π and can be adjustable by changing the strength of the quadrupoles in the matching section to compensate the emittance dilution due to CSR effect [14]. Finally, the DBA section, consisting of two 3-degree dipoles and a quadrupole inserted between them, bend the electron beam with a total angle of zero, making the two undulator lines parallel to each other. The scheme is designed and simulated based on Elegant code [15]. The dispersion function is optimized to be almost zero. The twiss functions in the switchyard for our scheme are shown in **Fig. 3**.

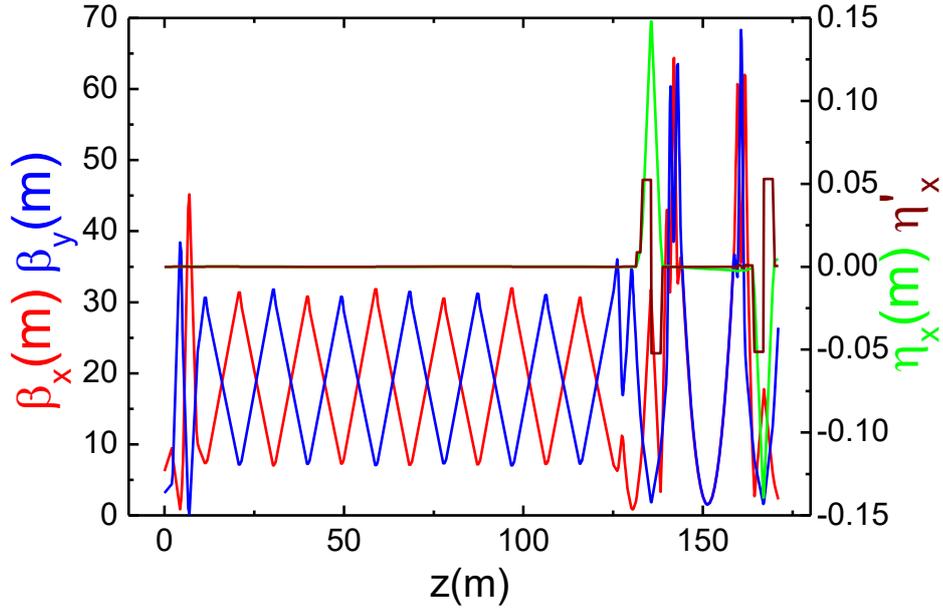

Fig. 3 Twiss functions in the switchyard

## 3 Beam collective effect
### 3.1 Resistive wall effects in beta matching section

As described in section 2, there is a beta matching section in between the main linac and the switchyard, which consists of 6 FODO cells to optimize the beta functions. The total length of the matching section is about 130 m and therefore the wakefield introduced by the resistive wall shall be taken into account. As we have known, the wakefield has both the transverse and the longitudinal components, the former introduces the beam a transverse kick and the latter brings energy spread to the beam. In the case of the single-pass linac, the longitudinal wakefield plays a more important role, therefore it needs us to choose the radius of the beam pipe appropriately to make the balance between the engineering cost and minimizing the energy spread introduced by the wakefield. As the results, **Fig. 4** shows the longitudinal wake behind a point charge in a round, metallic pipe with various radiuses [16], and **Fig. 5** illustrates us the central energy variation along the beam at the exit of the FODO cells for different beam pipe radiuses. In the figures, one can see that although the pipe with larger radius has less effect on the beam energy spread, the pipe with 17.5 mm radius converges well enough and make a good balance between the engineering considerations such as the design and the manufacture of the quadrupoles outside the pipe and the minimization of the energy spread introduced by the wakefield.

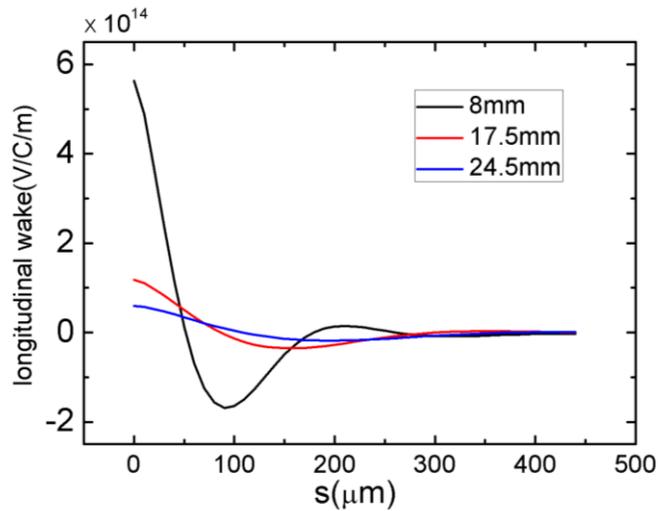

**Fig. 4**  Longitudinal wakefield behind a point charge in a round, metallic pipe with various radiuses

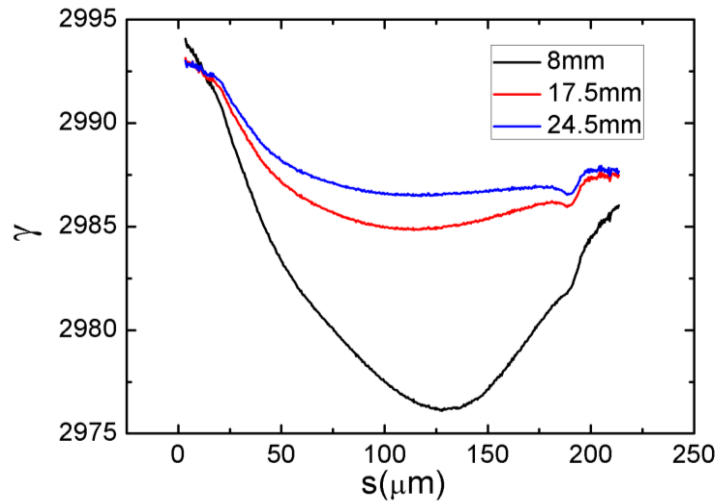

**Fig. 5** Variation of central energy along the beam for various pipe radiuses at the exit of FODO cells

**3.2  Microbunching instability**

Numerical and theoretical investigations of high-brightness electron beam demonstrate that the microbunching instability, driven by the coherent synchrotron radiation (CSR) effect, the longitudinal space charge (LSC) effect and the longitudinal wake field, may significantly degrade the beam quality [17, 18]. At the same time, the FODO section upstream of the switchyard is so long that even very small amount of density modulation converted from the chicane (BC2) may accumulate large amount of energy modulations, which leads to lots of fragments in the phase space. Three-dimensional start-to-end simulations have been carried out considering the CSR and the LSC effect, in order to have a research on the beam quality evaluation. The emittance evolution through the switchyard is shown in **Fig. 6**, from which one can find that the horizontal emittance growth is suppressed to about 8% and the vertical emittance maintains very well. The energy and the current distributions at the exit of the switchyard considering the microbunching instability is shown in **Fig. 6** as well. It is easy to find out that amount of energy modulation and density modulation are introduced because of the microbunching instability, which may degrade the FEL performance.

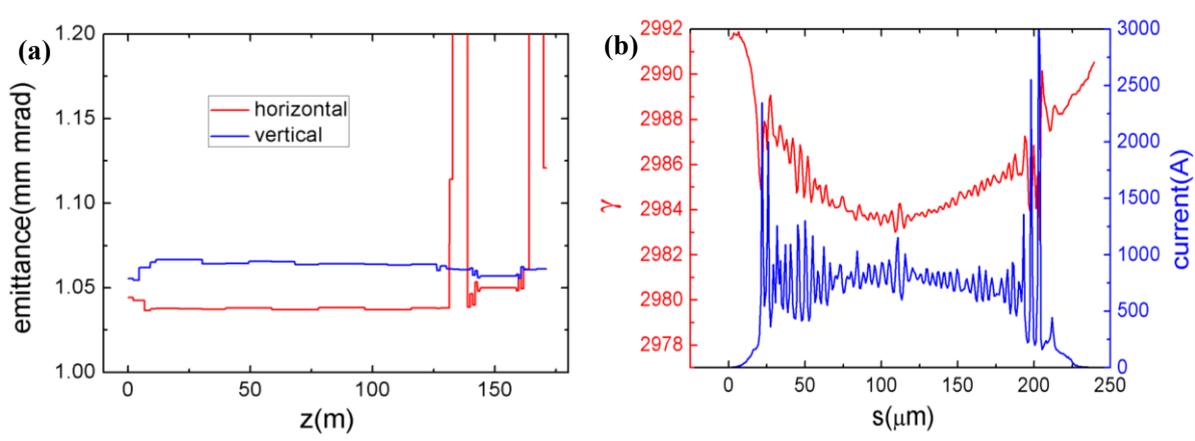

**Fig. 6** Emittance evolution through the switchyard (a) and energy and current distributions at the exit of the switchyard (b)

## 4  Research on jitters and misalignments

The FEL has a high requirement on the electron beam quality. The machine errors are arisen randomly in the real operation condition, which may lead to the jitters of the electron beam. In order to maintain the electron trajectory stability at the undulator entrance, we need to set the component tolerances and estimate the expected trajectory stability. The jitters may come from magnet current jitter, transverse vibration, misalignment and bunch length variations due to CSR. In our study, we use the dimensionless, normalized trajectory amplitude Ax/Ay rather than simple random error tracking [19], which can be expressed as:

$$A_x = \sqrt{\frac{x^2 + (x\alpha + x'\beta)^2}{\varepsilon\beta}} \quad (1)$$

Where x and x' are the horizontal centroid and angular betatron oscillation components. α, β, ε are the twiss parameters and geometric emittance at any point along the accelerator. Ay can be expressed similarly for vertical plane.

In order to show the final trajectory stability at the linac exit, we carried out the jitter studies on the whole linac including the switchyard. The component tolerances of the whole machine and the corresponding normalized trajectory amplitude are shown in **Table 2**. Meanwhile, the simulated trajectory jitters at the switchyard exit are shown in **Fig. 7**.

Table 2  Summary of trajectory jitters

| Mechanism | RMS Errors | Ax (%) | Ay (%) |
|---|---|---|---|
| Corrector Current | 5e-4 | 5 | 7 |
| Bend Current | 5e-5 | 2 | 0 |
| Quad vibration | 150 nm | 5 | 8 |
| Quad Current | 2e-4 | 5 | 4 |
| Quad Misalignment | 200 μm | | |
| Kicker | 5e-4 | 12 | 0 |
| Septum | 1e-5 | 4 | 0 |
| CSR + $\sigma_z$ jitter | 5% | 10 | 0 |
| Total (RMS) | | 18 | 11 |

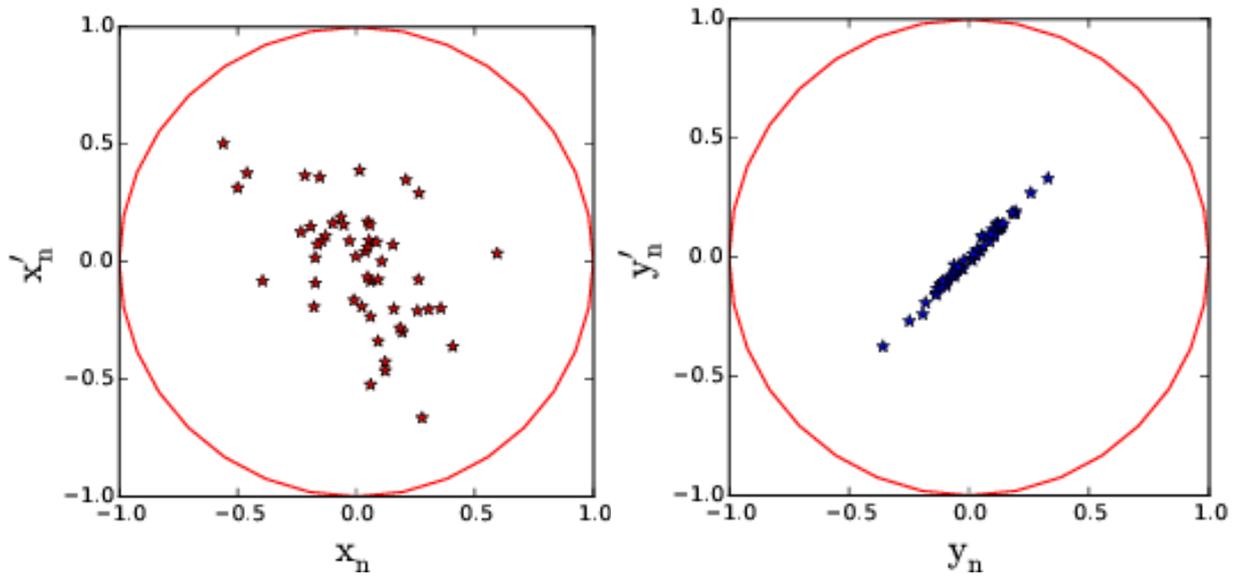

**Fig. 7**  Trajectory jitters at switchyard exit (50 random errors)

In addition, the orbit corrections of the electron beam in the switchyard, as shown in **Fig. 8**, have been researched as well. It is easy to find out that the maximum of the orbit deviation can be controlled within 100 μm.

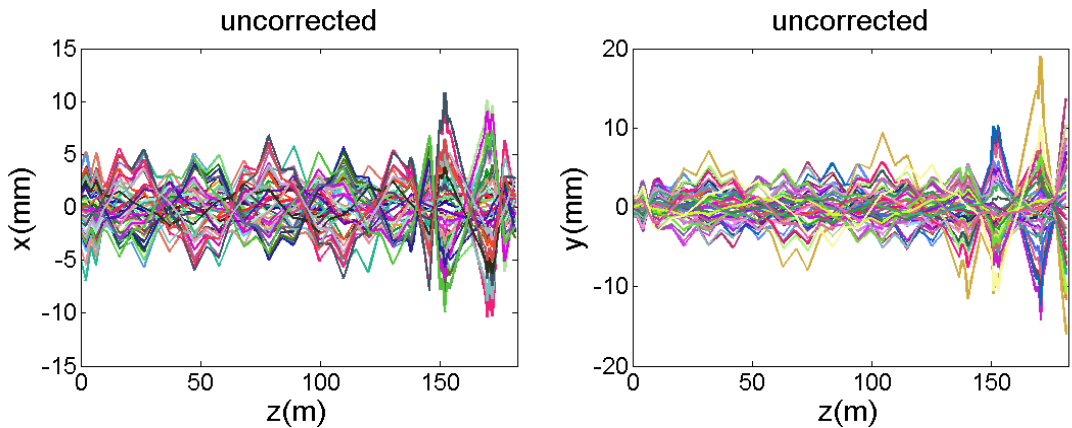

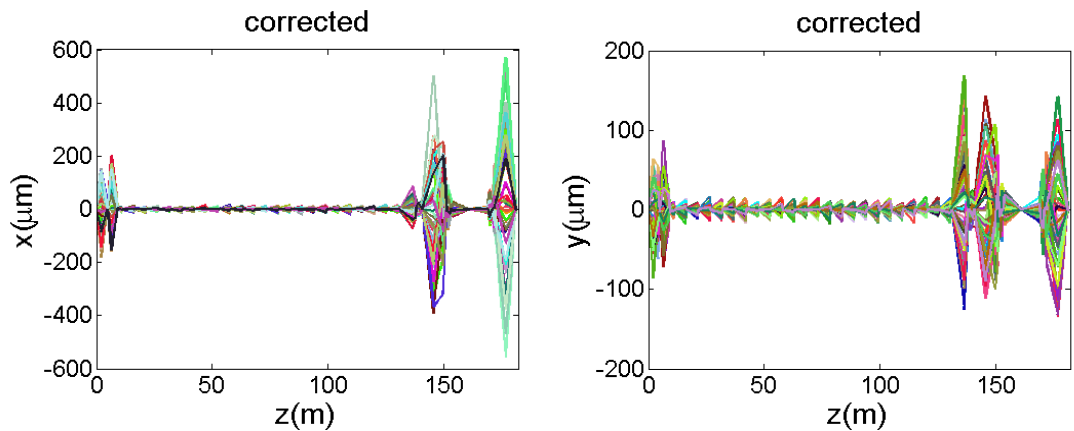

**Fig. 8**  The orbit deviation corrections of the electron beam in the switchyard

## 5  Conclusion

In this article, a feasible switchyard design is proposed for the Shanghai soft x-ray free-electron laser facility. Three-dimensional start-to-end simulations have been carried out to have a research on the beam quality evolutions as the beam passes through the switchyard. The simulation results show that the emittance of the beam at the exit of the switchyard increases less than 8%. The beam collective effects have been studies to research the beam quality evaluation. Finally, the longitudinal and the transverse jitters of the beam have been researched to optimize the lattice of the switchyard with achievable machine tolerance.